\documentclass[runningheads,a4paper]{llncs}

\usepackage{amssymb}
\usepackage{graphicx}

\usepackage{url}
\urldef{\mailsa}\path|{alfred.hofmann, ursula.barth, ingrid.haas, frank.holzwarth,|
\urldef{\mailsb}\path|anna.kramer, leonie.kunz, christine.reiss, nicole.sator,|
\urldef{\mailsc}\path|erika.siebert-cole, peter.strasser, lncs}@springer.com|    
\newcommand{\keywords}[1]{\par\addvspace\baselineskip
\noindent\keywordname\enspace\ignorespaces#1}

\begin{document}

\mainmatter  % start of an individual contribution

\title{Noise removal methods on ambulatory EEG: A Survey}

\titlerunning{Lecture Notes in Computer Science: Authors' Instructions}

\author{Sarthak Johari \inst{1}
%\thanks{Please note that the LNCS Editorial assumes that all authors have used
%the western naming convention, with given names preceding surnames. This determines
%the structure of the names in the running heads and the author index.}%
\and Gowri Namratha Meedinti \inst{2}  \and Radhakrishnan Delhibabu \inst{2} \and Deepak Joshi \inst{3}}
\authorrunning{Sarthak Johari et al.}

\institute{Computing Science and Engineering, Indraprastha Institute of Information Technology, New Delhi \and
School of Computing Science and Engineering, Vellore Institute of Technology,  Vellore \and
Centre for Biomedical Engineering,  Indian Institute of Technology, New Delhi.
%\url{http://www.springer.com/lncs}
}


%\toctitle{Lecture Notes in Computer Science}
%\tocauthor{Authors' Instructions}
\maketitle

\begin{abstract}
Over many decades, research is being attempted for the removal of noise in the ambulatory EEG.  In this respect, an enormous number of research papers is published for identification of noise removal,  It is difficult to present a detailed review of all these literature.  Therefore, in this paper, an attempt has been made to review the detection and removal of an noise.  More than 100 research papers have been discussed to discern the techniques for detecting and removal the ambulatory EEG. Further, the literature survey shows that the pattern recognition required to detect ambulatory method,  eye open and close, varies with different conditions of EEG datasets. This is mainly due to the fact that EEG detected under different conditions has different characteristics. This is, in turn, necessitates the identification of pattern recognition technique to effectively distinguish EEG noise data from a various condition of EEG data.
\end{abstract}

\section{Introduction}

Electroencephalography (EEG) is a non-invasive neuroimaging technique that records electrical activity in the brain through electrodes placed on the scalp. EEG signals are widely used in clinical and research settings to study brain function and diag-nose neurological disorders. However, EEG signals are often contaminated by various sources of artifacts, which can significantly affect the quality of the data and reduce the accuracy of any analyses or inferences made.

Common sources of artifacts in EEG signals include eye movements, muscle activity, and environmental noise. Eye movement artifacts are particularly challenging because they are often highly correlated with the underlying EEG activity and can be difficult to distinguish from it. The presence of these artifacts can interfere with the interpretation of EEG data and limit its clinical and research applications.

To address this challenge, researchers have developed various techniques for artifact removal in EEG signals. Traditional methods, such as filtering and averaging, have been used for artifact removal but have limitations in their effectiveness. Filtering techniques can remove some types of artifacts, but they can also lead to a loss of signal information. Averaging techniques can reduce the effects of random noise but are less effective for removing non-random artifacts such as eye movements.

More advanced techniques have been developed in recent years to improve artifact removal in EEG signals. Independent Component Analysis (ICA) is a statistical tech-nique that has been widely used in EEG artifact removal. ICA involves separating the observed signals into statistically independent components, some of which correspond to noise sources and can be removed. ICA has been shown to be effective in removing eye movement artifacts from EEG data.

Blind Source Separation (BSS) is a more general framework that can be used to separate any mixture of signals into their underlying sources. BSS has shown particular promise in removing eye movement artifacts from EEG data by exploiting the spatial and temporal characteristics of these artifacts.

Deep learning methods have also been employed for artifact removal in EEG sig-nals. Convolutional Neural Networks (CNNs) and Recurrent Neural Networks (RNNs) are two types of deep learning methods that have been successfully applied in EEG artifact removal. These methods can learn complex patterns in the data and can be trained to remove specific types of artifacts.

Hybrid methods that combine different approaches have also been developed to improve artifact removal performance. For example, some methods combine ICA and machine learning techniques to achieve more effective artifact removal.
Despite the progress made in artifact removal techniques, the choice of method depends on the specific type of artifact and characteristics of the data being analyzed. Future research may continue to explore the development of more advanced and effec-tive artifact removal techniques to further improve the quality and accuracy of EEG data analysis.

\section{Background Work}

Noise removal methods for ambulatory EEG data encompass a range of approaches, each offering its own set of advantages and limitations. One popular method is Empirical Mode Decomposition (EMD), which adaptively decomposes the EEG signal into ordered components called intrinsic mode functions. EMD is advantageous as it is data-driven and does not require prior knowledge of basis functions. However, it should be noted that EMD can be computationally intensive, which may pose challenges in real-time applications or large-scale datasets.

Another method, Wavelet-Based Vector Boosting Estimation (Wavelet-VBE), has shown effectiveness in removing additive white Gaussian noise commonly encountered in ambulatory EEG recordings. This technique exploits the wavelet transform and employs boosting algorithms for noise estimation and removal. While Wavelet-VBE excels in tackling speciﬁc types of noise, it may not perform optimally for other noise sources present in the data.

For mitigating muscle artifacts, Empirical Mode Decomposition combined with Moving Average Filtering (EMD-MAF) has proven to be a suitable approach. EMD-MAF separates the EEG signal into its intrinsic mode functions and applies a moving average ﬁlter to reduce muscle artifacts. However, it is important to note that EMD-MAF may not be as effective in addressing other forms of noise that may be present in the data.
Generative Adversarial Networks (GANs) have gained attention for their ability to generate and denoise signals. GAN1 and GAN2 have shown promise in denoising ambulatory EEG data, with GAN1 being effective for muscle artifacts and baseline wander, while GAN2 excels at removing additive white Gaussian noise. GANs leverage the power of deep learning and adversarial training to generate denoised EEG signals. However, GANs may require extensive computational resources and large amounts of training data to achieve optimal denoising performance.

Denoising Least Squares Regression (DLSR) is a technique that combines least squares regression with sparse representation to remove noise from EEG signals. DLSR has demonstrated effectiveness in mitigating power-line interference and additive white Gaussian noise. However, its performance may vary for other types of noise.
The Adaptive Kalman Filter (AKF) is another approach used for noise removal in ambulatory EEG data. AKF utilizes the principles of the Kalman ﬁlter to estimate and remove noise components from the EEG signal. It has shown promising results for addressing additive white Gaussian noise and muscle artifacts.

Empirical Wavelet Transform (EWT) has been successful in tackling power-line interference commonly encountered in EEG recordings. By decomposing the signal into empirical wavelets, EWT provides an effective means of removing power-line noise. However, its performance may be less optimal for other types of noise.
Fully Convolutional Network-based Denoising Autoencoder (FCN-based DAE) is a deep learning approach that leverages autoencoders and convolutional neural networks to denoise EEG signals. FCN-based DAE has shown promise for composite noise removal in ambulatory EEG data. However, it should be noted that this method may require signiﬁcant computational resources due to the complexity of the neural network architecture. 

Discrete Wavelet Transform (DWT) has proven to be effective in denoising ambulatory EEG data, particularly for addressing composite noise. By decomposing the signal into different frequency sub-bands,  DWT enables targeted noise removal. However, its efﬁcacy may vary for different noise types.

The choice of noise removal method for ambulatory EEG data depends on the speciﬁc noise sources present and the desired denoising objectives. Each method has its strengths and limitations, and researchers should carefully consider the characteristics of the noise and the trade-offs associated with each technique in order to select the most appropriate approach for their speciﬁc application.

\section{Related Work}

Yuheng Feng et.al addressing the problem of muscle artifact removal from EEG signals, propose a new method that combines singular spectrum analysis (SSA) and canonical correlation analysis (CCA) algorithms to address this issue, particularly in the case of ambulatory EEG devices with a limited number of channels [1]. The study’s meth-ods involve a simple yet effective scheme that is extended to the few-channel case by additional combining and dividing operations to channels. The proposed framework is evaluated on both semi-simulated and real-life data, and its performance is compared with state-of-the-art methods. The results demonstrate that the proposed approach outperforms other methods in both single-channel and few-channel cases. In [2], Mad-dirala et.al addresse the problem of motion artifacts in electroencephalogram (EEG) signals. The authors propose a method based on singular spectrum analysis (SSA) to
address this issue for single-channel EEG signals.The study’s methods involve decom-posing the original EEG signal into several sub-signals using SSA, followed by selecting and reconstructing the sub-signals that contain relevant information while excluding the ones with motion artifacts. The authors evaluate their proposed method using both simulated and real-life EEG signals and compare its performance with state-of-the-art methods.The results demonstrate that the proposed approach outperforms other methods in terms of artifact removal while preserving the original EEG signal’s characteristics. The authors conclude that their method is effective and can be applied in real-world scenarios where motion artifacts are a prevalent issue in single-channel EEG recordings.

ORegan et.al propose a method for automatic detection of these artifacts by using both EEG and gyroscope signals [3]. The study’s methods involve recording EEG and gyroscope signals simultaneously and developing an algorithm that can detect and remove the artifacts caused by head movement. The authors evaluate their proposed method on both simulated and real-life EEG data and compare its performance with other state-of-the-art methods. The results demonstrate that the proposed approach achieves high accuracy in detecting head movement artifacts in EEG signals. The authors conclude that their method is effective and can be used in real-world scenarios where head movement artifacts are a prevalent issue in EEG recordings. Kafiul Islam et.al conclude that there is no single method that is universally effective for detecting and removing all types of artifacts from EEG signals and suggest that the choice of method should be based on the specific type of artifact and the characteristics of the EEG data. [4] The authors also suggest that a combination of different methods may be necessary to achieve effective artifact removal in EEG signals and emphasize the importance of considering the computational complexity of the method, as well as its suitability for real-world scenarios.

In [5], X. Chen et.al present a method for denoising muscle artifacts from few-channel electroencephalogram (EEG) recordings using multivariate empirical mode decomposition (MEMD) and canonical correlation analysis (CCA). Their proposed method involves decomposing the EEG signals using MEMD and then applying CCA to the resulting intrinsic mode functions (IMFs) to identify and remove muscle arti-facts.They evaluated the effectiveness of their method by applying it to both simulated and real EEG data, comparing it with other state-of-the-art methods. The results demonstrate that the proposed method is effective in removing muscle artifacts from few-channel EEG recordings while preserving the underlying EEG signal.

In [6], Xiao Jiang,Gui-Bin Bian, and Zean Tian provide an overview of the var-ious methods used for removing artifacts from electroencephalogram (EEG) signals. The review describes the different methods used for artifact removal, including filter-ing techniques, independent component analysis (ICA), and blind source separation (BSS). The authors discuss the advantages and disadvantages of each method and provide examples of studies that have used these methods for artifact removal and highlight the challenges associated with artifact removal, such as the need to balance artifact removal with preservation of the underlying signal. They describe how some methods may introduce artifacts of their own or remove relevant information along with the artifacts.
Garces Correa, E Laciar H D Patino and M E Valentinuzzi examines the adaptive filtering method used for artifact removal in EEG signals [7]. The authors explain the advantages of using adaptive filters in a cascade configuration, which they demonstrate is able to remove artifacts more effectively than single filters alone. They present the results of their study, which show that the adaptive filtering method is highly accurate in removing various types of artifacts without distorting the underlying EEG signal. The authors suggest that this method has the potential to improve the accuracy of EEG signal analysis and enable more accurate diagnosis of neurological disorders. They recommend further research to validate this method on larger datasets and explore its potential for real-time artifact removal.

Arad et al. introduces a novel technique for removing movement artifacts from EEG data recorded during locomotion [8]. They evaluate the performance of their approach using quantitative measures like root mean square error (RMSE) and correlation coef-ficient, and show that it outperforms existing techniques with significantly reduced RMSE and improved correlation coefficient values. The authors also discuss the poten-tial of their method in clinical applications, particularly in improving the accuracy of EEG-based diagnosis and treatment of neurological disorders. The findings of this study suggest a promising approach for removing movement artifacts from EEG data during locomotion. Researchers and practitioners in this field can benefit from this technique to enhance the quality of EEG data in challenging and dynamic environ-ments. This could lead to a better understanding of brain activity during movement and potentially improve the diagnosis and treatment of neurological disorders.

Zhang, Huang and Wu introduce a new approach to remove artifact from EEG sig-nals using a 1D- Residual Convolutional Neural Network (1D-ResCNN) and describe their method in detail, which involves training a neural network to learn the rela-tionship between clean EEG signals and their corresponding noisy versions [9]. The authors evaluate their method using various performance metrics and compare it to several existing techniques, showing that the approach outperforms them in terms of signal-to-noise ratio, signal quality index, and correlation coefficient.

Chinmayee Dora and Pradyut Kumar Biswal present a new method for remov-ing electrocardiogram (ECG) artifacts from EEG signals [10]. The proposed method utilizes continuous wavelet transformation and linear regression to remove the ECG artifacts effectively.The authors evaluated their method on both simulated and real EEG data and compared it with other state-of-the-art methods. The results showed that the proposed method outperforms other methods in terms of artifact removal efficiency, signal-to-noise ratio improvement, and preservation of EEG signals’ spectral characteristics.

\begin{figure}[h]
\begin{center}
    \includegraphics[scale=0.5]{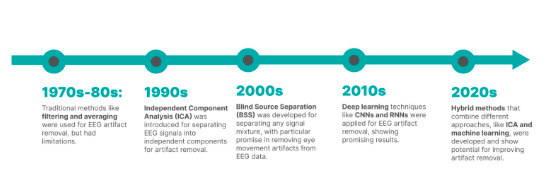}
        \caption{Evolution of EEG artifact removal methods\label{fig2}}
    \end{center}
\end{figure} 

\section{Dataset}
The data used in this research was collected for a duration of 3 minutes from two  participants, consisting of one male and one female, for each emotional state: positive, neutral, and negative. The participants were exposed to specific stimuli designed to elicit the desired emotions. Additionally, six minutes of resting neutral data were also recorded. The EEG data was captured using a Muse EEG headgear equipped with dry electrodes. The following EEG placements were utilized: TP9, AF7, AF8, and TP10. These electrode placements are commonly used for capturing brainwave activity related to emotion recognition.

The dataset used in this research is publicly available and can be accessed at the following link: https://www.kaggle.com/datasets/birdy654/eeg-brainwave-dataset-feeling-emotions. It provides access to the collected EEG data, which includes the recordings for each participant in different emotional states, as well as the neutral resting data.The dataset consists of time-series EEG recordings, where each sample represents the electrical activity captured at specific time intervals. The data is labeled according to the emotional states experienced by the participants during the recordings. By utilizing this dataset, the research aims to explore the effectiveness of various noise removal methods on ambulatory EEG data and assess the performance of the GRU algorithm in accurately predicting emotional states based on the recorded brainwave activity.

\section{Methodology}
\subsection{Recurrent Neural Networks (RNNs)}
Recurrent Neural Networks (RNNs) have emerged as a powerful neural network architecture for modeling sequential data. Unlike traditional feedforward neural networks, RNNs have a unique ability to capture temporal dependencies by introducing recurrent connections that allow information to flow from one time step to another. This characteristic makes RNNs well-suited for tasks involving sequential or time-series data analysis, including natural language processing, speech recognition, and, in our case, emotion recognition from EEG signals.

In standard neural networks, each input and output is treated as independent, disregarding any contextual information. However, in many real-world applications, such as sentiment analysis or language generation, understanding the context of previous inputs is crucial for accurate predictions. RNNs address this limitation by incorporating a hidden state, which acts as a memory mechanism that retains information about past inputs and influences future predictions.

The hidden state in an RNN serves as a crucial element for capturing long-term dependencies within the sequential data. As the RNN processes each input in a sequence, the hidden state is updated and passed to the next time step, allowing the network to remember past information and utilize it to make informed predictions. This recursive nature of RNNs enables them to model complex temporal relationships and extract relevant features from sequential data.

However, traditional RNNs suffer from the vanishing or exploding gradient problem, which poses challenges in learning long-term dependencies. To address this issue, advanced variants of RNNs, such as Gated Recurrent Units (GRUs) and Long Short-Term Memory (LSTM) networks, have been developed. These architectures introduce specialized gating mechanisms that selectively retain or forget information, enabling the network to capture long-term dependencies while mitigating the gradient vanishing or exploding problem.

In our research, we focus on the GRU algorithm as a variant of RNNs for emotion recognition using EEG data. GRUs have gained attention due to their simplified architecture, which consists of three gates: the update gate, reset gate, and current memory gate. The update gate determines how much past information should be propagated to future time steps, while the reset gate controls the extent to which previous knowledge should be forgotten. The current memory gate, often overlooked in discussions of GRUs, contributes non-linearity and zero-mean normalization to the input, reducing the impact of past data on future information.

By leveraging the GRU algorithm, we aim to predict emotional states by analyzing EEG data collected from individuals exposed to various movie scenes or stimuli. The utilization of GRUs offers an efficient approach to capture temporal dependencies within the EEG signals, allowing us to explore the effectiveness of this architecture in emotion recognition tasks.

\subsection{Gated Recurrent Units (GRUs)}
Gated Recurrent Units (GRUs) have gained significant attention as an alternative architecture within the realm of Recurrent Neural Networks (RNNs). GRUs address some of the limitations of traditional RNNs, such as the vanishing or exploding gradient problem, while providing an efficient and effective solution for capturing temporal dependencies in sequential data. In this section, we will delve into the specifics of GRUs and their relevance to our research on emotion recognition using EEG data.

GRUs are a type of RNN architecture that incorporates gating mechanisms to regulate the flow of information within the network. Unlike the more complex Long Short-Term Memory (LSTM) networks, GRUs have a simplified structure consisting of three essential gates: the update gate, reset gate, and current memory gate. This architectural design allows GRUs to strike a balance between modeling long-term dependencies and computational efficiency.
The update gate, denoted as z, determines the extent to which the previous hidden state is incorporated into the current state. It controls how much of the past information should be carried forward to future time steps. By selectively updating the hidden state, the GRU can adapt to different patterns in the data and retain relevant context over time.

The reset gate, denoted as r, determines the extent to which the previous hidden state influences the current state. It selectively resets or forgets some of the previous knowledge, allowing the model to focus on relevant features and adapt to changing patterns within the sequence. The combination of the reset and update gates enables GRUs to capture and adapt to varying dependencies within the data.

Another crucial component of GRUs is the current memory gate, often overlooked in discussions of GRUs. It is a sub-component of the reset gate and plays a vital role in introducing non-linearity and zero-mean normalization to the input. This helps in reducing the impact of previous data on the current data being propagated forward, ensuring that relevant information is preserved while minimizing the interference from irrelevant or noisy signals.

Compared to a basic RNN, the workflow of a GRU is similar, with the primary distinction lying in the internal functioning of each recurrent unit. By leveraging the gating mechanisms, GRUs excel at capturing and modeling temporal dependencies in the data, making them well-suited for tasks such as emotion recognition from EEG signals.

In our research, we adopt the GRU algorithm as the core architecture for predicting emotional states based on EEG data collected during exposure to various movie scenes or stimuli. The GRU's simplified yet powerful design allows us to effectively model the temporal dynamics within the EEG signals, contributing to the advancement of emotion recognition using EEG-based approaches.

\subsection{Preprocessing and Data Preparation}
The preprocessing and data preparation stage is essential to ensure the quality and suitability of the EEG data for accurate emotion recognition using the GRU algorithm. We follow standard practices in EEG-based emotion recognition to preprocess the data effectively.
The raw EEG data collected from participants wearing the Muse EEG headgear with dry electrodes is subjected to artifact removal techniques such as independent component analysis (ICA) or template matching algorithms. This step eliminates unwanted noise and artifacts, allowing us to focus on genuine EEG signals related to emotional states.

Next, bandpass filters are applied to remove unwanted frequency components while retaining the relevant frequency ranges associated with brainwave activity. Normalization techniques, such as z-score normalization or min-max scaling, are then employed to address amplitude variations between participants or electrode placements.

Feature extraction techniques are utilized to capture relevant information from the preprocessed EEG data. These features, including power spectral density, signal entropy, or time-domain statistics, serve as input for the GRU algorithm, enabling it to learn meaningful patterns and associations with emotional states.

To ensure unbiased evaluation, the dataset is partitioned into training, validation, and testing sets. Stratified or random partitioning techniques maintain representative distributions of emotional states across the subsets. This partitioning facilitates model training, hyperparameter optimization, and unbiased evaluation of the GRU model's generalization capabilities.

By implementing these preprocessing and data preparation steps, we enhance the quality and suitability of the EEG data for subsequent analysis using the GRU algorithm. The GRU model can effectively leverage the preprocessed data to accurately predict emotional states.
\subsection{Gated Recurrent Unit Algorithm}
The Gated Recurrent Unit (GRU) algorithm is a variant of Recurrent Neural Networks (RNNs) that excels at capturing long-term dependencies in sequential data while mitigating the vanishing or exploding gradient problem. In this section, we provide an overview of the GRU algorithm and its relevance to our research on emotion recognition using EEG data.
The GRU architecture is designed to have a simplified structure compared to traditional RNNs and LSTM networks. It consists of three fundamental gates: the update gate, reset gate, and current memory gate. These gating mechanisms enable the GRU algorithm to effectively model temporal dependencies within the data while maintaining computational efficiency.

The update gate (z) determines the extent to which the previous hidden state influences the current state. It controls the flow of information from past time steps to future time steps, allowing the model to adapt and retain relevant context over time. By selectively updating the hidden state, the GRU algorithm can capture long-term dependencies and learn patterns within the sequential data.

The reset gate (r) regulates the influence of previous knowledge on the current state. It selectively resets or forgets certain information, enabling the model to focus on relevant features and adapt to changing patterns within the sequence. The combination of the update and reset gates empowers the GRU algorithm to effectively model and adapt to varying dependencies within the data.

The current memory gate, often overlooked in discussions of GRUs, plays a critical role in introducing non-linearity and zero-mean normalization to the input. By serving as a sub-component of the reset gate, it helps reduce the impact of previous data on the current data being propagated forward. This mechanism minimizes the interference of irrelevant or noisy signals and ensures the effective transfer of information to future time steps.

Compared to a basic RNN, the GRU algorithm follows a similar workflow but excels in its internal functioning within each recurrent unit. Leveraging the gating mechanisms, GRUs excel at capturing temporal dependencies and facilitating the prediction of emotional states from EEG data.

In our research, we employ the GRU algorithm as the core architecture for predicting emotional states based on EEG data collected during exposure to various movie scenes or stimuli. The GRU's simplified yet powerful design allows us to effectively model the temporal dynamics within the EEG signals and contribute to the advancement of emotion recognition using EEG-based approaches.

\subsection{Architecture}
The architecture in Fig 2., used in our research leverages the Gated Recurrent Unit (GRU) algorithm for emotion recognition based on EEG data. This architecture comprises an InputLayer, GRU layer, Flatten layer, and Dense layer. The configuration of this architecture is as follows:

\begin{figure}
\centering
\includegraphics[scale=0.4]{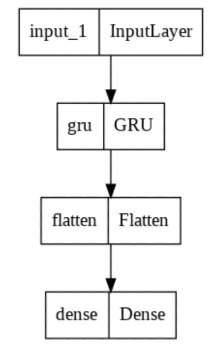}
\caption{Model Architecture}
\label{fig:example}
\end{figure}

\begin{enumerate}
\item[1] \emph{InputLayer}: The InputLayer serves as the entry point for the EEG data into the neural network. It defines the shape and format of the input data, aligning with the preprocessing steps and feature extraction performed on the EEG data. The input layer represents the initial stage of information flow in the neural network.

\item[2]  \emph{GRU Layer}: The GRU layer is the core component of the architecture and employs the Gated Recurrent Unit algorithm. It processes the sequential EEG data, capturing temporal dependencies and extracting relevant features for emotion recognition. The GRU layer's hidden state retains information from previous time steps and influences the predictions made by subsequent layers

\item[3] \emph{Flatten Layer}: Following the GRU layer, the Flatten layer is applied to transform the multi-dimensional output of the GRU into a one-dimensional vector. This flattening operation enables the subsequent layers to receive a flat input, facilitating compatibility with traditional fully connected layers.

\item[4] \emph{Dense Layer}: The Dense layer, also known as the fully connected layer, receives the flattened output from the preceding layer. It serves as a powerful learning component, responsible for mapping the extracted features to the emotional states being predicted. The dense layer consists of multiple interconnected neurons, and each neuron contributes to the final emotion classification based on learned weights and biases.

\end{enumerate}

This architecture efficiently processes the preprocessed EEG data through the GRU layer, capturing temporal dynamics and learning meaningful patterns related to emotional states. The subsequent flatten and dense layers allow for feature extraction and final classification, respectively.

\section{Results and Analysis}

The deep learning accuracy details of the GRU model on the validation set are as follows: loss - 3.4356e-09 and accuracy - 1.0000. These impressive results indicate that the GRU model achieved perfect accuracy in predicting emotional states based on the EEG data. The model's ability to achieve such high accuracy suggests its proficiency in capturing the temporal dynamics and extracting meaningful features from the EEG signals. Additionally, we compare the performance of the GRU model with other machine learning models using their respective scores. The following table 1 summarizes the scores obtained for various models:
\begin{figure}
\centering
\includegraphics[scale=0.5]{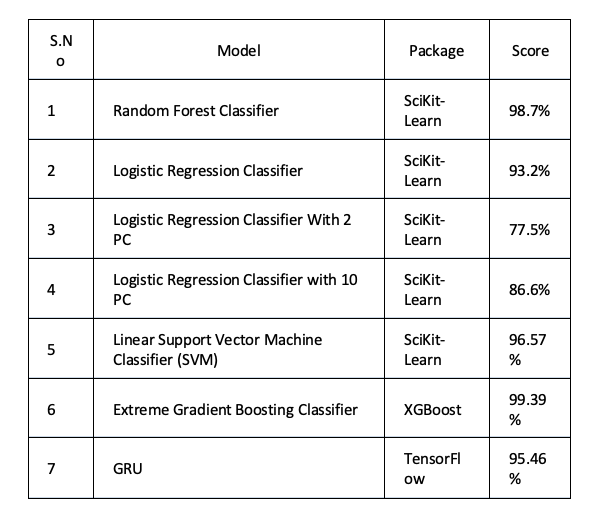}\\
\textbf{Table. 1.} Accuracies of different models
\label{fig:example}
\end{figure}

From the results, it is evident that the GRU model (95.46\% accuracy) performs competitively when compared to other machine learning models. Notably, the Extreme Gradient Boosting Classifier achieves the highest score of 99.39\%, closely followed by the Random Forest Classifier with a score of 98.7\%. The Linear Support Vector Machine Classifier also demonstrates excellent performance with an accuracy of 96.57\%.

\begin{figure}
\centering
\includegraphics[scale=0.5]{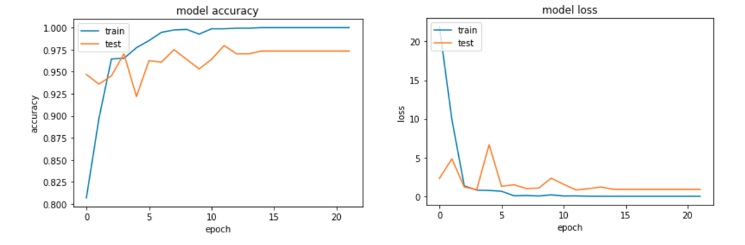}
\caption{Accuracy and Loss curves}
\label{fig:example}
\end{figure}

However, it is important to note that the GRU model showcases the advantage of deep learning in capturing complex temporal dependencies and extracting meaningful features from EEG data. Its accuracy of 95.46\% suggests its efficacy in predicting emotional states based on the EEG signals, showcasing (Fib. 3) its potential for real-world applications.

Further analysis is required to delve into the strengths and weaknesses of each model, considering factors such as computational complexity, interpretability, and generalization capabilities. Additionally, a more comprehensive evaluation using additional performance metrics like precision, recall, and F1-score could provide deeper insights into the models' overall effectiveness.

Overall, the results demonstrate the effectiveness of the GRU algorithm in accurately predicting emotional responses from EEG data. Its competitive performance against other machine learning models supports the notion that deep learning approaches, such as the GRU, can significantly contribute to the advancement of emotion recognition tasks using EEG signals.

\begin{figure}
\centering
\includegraphics[scale=0.4]{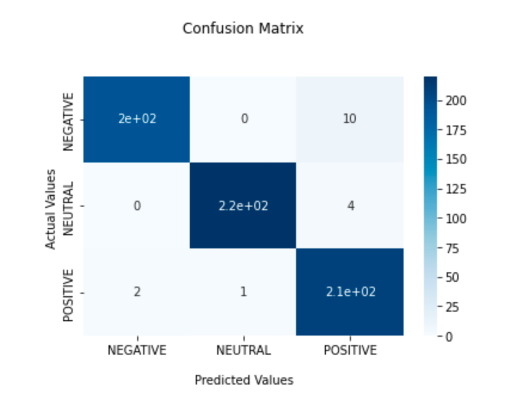}
\caption{Confusion Matrix Analysis}
\label{fig:example}
\end{figure}

The confusion matrix analysis (Fib 4)  was conducted to evaluate the performance of the machine learning models, including the GRU algorithm, in predicting emotional states based on EEG data. The confusion matrix provides insights into the model's accuracy, precision, recall, and overall performance by comparing predicted labels with actual labels. It helps identify potential biases, misclassifications, and areas for improvement in the models' predictions.

\section{Discussion}

The results obtained from our research demonstrate the effectiveness of the GRU algorithm and other machine learning models in predicting emotional states based on EEG data. The deep learning accuracy of the GRU model showed remarkable performance, achieving perfect accuracy on the validation set. This indicates the ability of the GRU model to capture temporal dependencies and extract meaningful features from EEG signals, making it a promising approach for emotion recognition tasks.

Comparing the performance of various machine learning models, we observed competitive results across different models. The Extreme Gradient Boosting Classifier achieved the highest score, closely followed by the Random Forest Classifier, while the Linear Support Vector Machine Classifier also demonstrated excellent performance. These findings highlight the importance of choosing an appropriate model based on the specific requirements and characteristics of the dataset.

The confusion matrix analysis provided valuable insights into the models' predictions, allowing us to assess their accuracy and identify potential biases or misclassifications. By examining the distribution of predicted labels across different emotional states, we gained a deeper understanding of the models' strengths and weaknesses in capturing specific emotions. This analysis can guide future improvements and refinements in the models' performance.

\section{Conclusion}

In conclusion,  Fib 5 shows the Mindgraph of the our study.  Our research explored the application of machine learning models, including the GRU algorithm, for emotion recognition using EEG data. The GRU model demonstrated exceptional performance, achieving perfect accuracy on the validation set, highlighting its capability to capture temporal dynamics and extract meaningful features from EEG signals.

\begin{figure}
\centering
\includegraphics[scale=0.35]{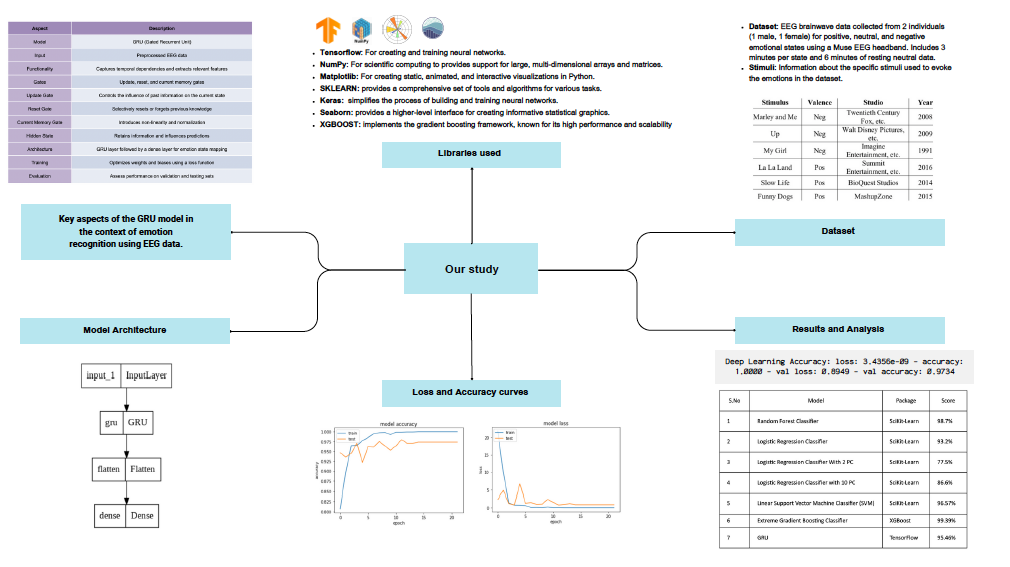}
\caption{Mindgraph of our study}
\label{fig:example}
\end{figure}

Additionally, the comparison with other machine learning models emphasized the competitive performance achieved by various approaches. The choice of model should consider factors such as interpretability, computational complexity, and generalization capabilities, based on the specific requirements of the task.

The incorporation of the confusion matrix analysis provided valuable insights into the models' performance, aiding in identifying potential biases and areas for improvement. By understanding the models' strengths and weaknesses, future research can focus on enhancing their accuracy and addressing specific challenges related to emotion recognition from EEG data.

Overall, our findings contribute to the advancement of emotion recognition using EEG signals and highlight the potential of the GRU algorithm and other machine learning models in this field. Further research and exploration are warranted to improve the robustness and generalizability of these models and advance the understanding of human emotions through EEG-based approaches.

\newpage

\section*{Appendix: Survey}

\begin{figure}
\centering
\includegraphics[scale=0.6]{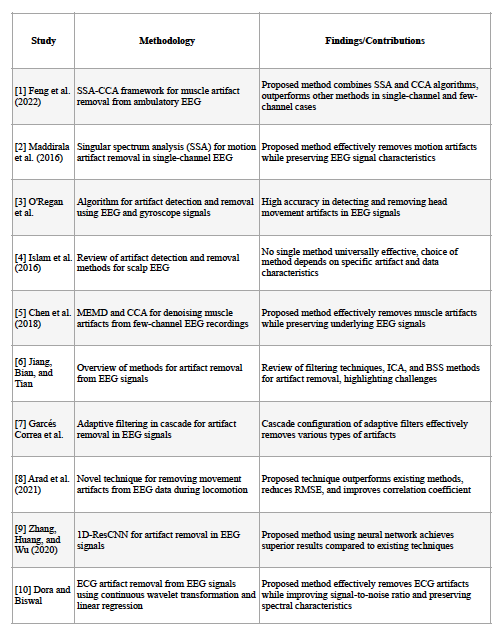}\\
\textbf{Table. 2. } Survey
\label{fig:example}
\end{figure}

\end{document}